\documentclass[twocolumn,prl,aps,nobalancelastpage,amssymb]{revtex4}

\usepackage{graphicx}

\def \LSCO{La$_{2-x}$Sr$_x$CuO$_4$}
\def \LSCOOD{La$_{1.7}$Sr$_{0.3}$CuO$_4$}
\def \beqn {\begin{equation}}
\def \bfig {\begin{figure}}
\def \btab {\begin{table}}
\def \eeqn {\end{equation}}
\def \efig {\end{figure}}
\def \etab {\end{table}}

\def \RH{$R_{\rm H}$}

\def \MRa{$\Delta \rho_{ab}/\rho_{ab}$}

\begin{document}
\title{Violation of the isotropic-$\ell$ approximation in overdoped \LSCO}

\author{A. Narduzzo$^1$, G. Albert$^1$, M. M. J. French$^1$, N. Mangkorntong$^2$, M. Nohara$^2$, H. Takagi$^2$ and N. E. Hussey$^1$}

\affiliation{$^1$H. H. Wills Physics Laboratory, University of Bristol, Tyndall Avenue, BS8 1TL, U.K.}
\affiliation{$^2$Department of Advanced Materials Science, Graduate School of Frontier Science, University of Tokyo,
Kashiwa-no-ha 5-1-5, Kashiwa-shi, Chiba 277-8651, Japan.}

\date{\today}

\begin{abstract}
Magnetotransport measurements on the overdoped cuprate \LSCOOD~are fitted using the Ong construction and band
parameters inferred from angle-resolved photoemission. Within a band picture, the low temperature Hall data can only be
fitted satisfactorily by invoking strong basal-plane anisotropy in the mean-free-path $\ell$. This violation of the
isotropic-$\ell$ approximation supports a picture of dominant small-angle elastic scattering in cuprates due to
out-of-plane substitutional disorder. We show that both band anisotropy and anisotropy in the elastic scattering
channel strongly renormalize the Hall coefficient in overdoped La$_{2-x}$Sr$_x$CuO$_4$ over a wide doping and
temperature range.
\end{abstract}

\maketitle

The normal state transport properties of cuprates \cite{HusseyReview} have proved as enigmatic as their high
temperature superconductivity. One of the most striking properties to be uncovered is the anomalously strong
temperature $T$- and doping $p$-dependence of the Hall coefficient \RH. This behavior is epitomized in
La$_{2-x}$Sr$_x$CuO$_4$ (LSCO), which spans the entire (hole-doped) phase diagram from the Mott insulator to the
non-superconducting metal.

Experimentally, $R_{\rm H}$ in LSCO is found to vary approximately as 1/$T$ over a wide temperature range
\cite{Nishikawa94}, in violation of standard Fermi-liquid theory. For $x < 0.05$, $R_{\rm H}$ scales roughly as 1/$x$
\cite{Takagi89, Ando04} but at higher doping levels, $R_{\rm H}$ falls more rapidly (by 2 orders of magnitude for $0.05
< x < 0.25$ \cite{Ong87, Takagi89, Ando04}) reflecting a crossover from a small to a large Fermi surface (FS). Finally
for $x > 0.28$, \RH($T$) is negative at elevated $T$ \cite{Hwang94, Tsukada06}.

Recent results from angle-resolved photoemission spectroscopy (ARPES) \cite{Ino02, Yoshida03, Yoshida06} appear to
support the doping evolution of the FS in LSCO inferred from \RH($T,x$). At low $x$, a weak quasiparticle peak appears
along the zone diagonal, forming an \lq arc' whose intensity increases with increasing $x$ \cite{Yoshida03} consistent
with the variation of the carrier number $n$. Beyond optimal doping, spectral intensity is observed everywhere on the
FS \cite{Yoshida06} and the Fermi level $\epsilon_F$ moves below the saddle near ($\pi$, 0), leading to a topological
shift from a hole-like, to an electron-like FS \cite{Ino02}.

Despite this apparent consistency between transport and spectroscopic probes, several outstanding issues remain, in
particular the magnitude and sign of $R_{\rm H}$ at low $T$. According to ARPES, the FS in LSCO is electron-like for $x
> 0.18$ \cite{Yoshida06}, yet \RH($T$$\rightarrow$0) remains positive up to $x$ = 0.34 \cite{Hwang94, Tsukada06}. This is
troublesome since at low $T$, where impurity scattering dominates, the mean-free-path $\ell$ is expected to become
independent of momentum {\bf k}. In this isotropic-$\ell$ regime, \RH(0) for a two-dimensional (2D) single-band metal
must reflect the sign of the dominant carrier, even if there are electron- and hole-like regions of FS \cite{Ong91},
with a magnitude equal to 1/$ne$.

Here we show that these discrepancies can be reconciled through a combination of strong band anisotropy and basal-plane
anisotropy in the $T=0$ scattering rate $\Gamma_0(\varphi)$. Though our analysis focuses on heavily overdoped
non-superconducting \LSCOOD~(LSCO30), our results are applicable over a wide range of $x$. By combining the Ong
construction for the Hall conductivity $\sigma_{xy}$ \cite{Ong91} with the 2D tight-binding band dispersion inferred
from ARPES \cite{Yoshida06}, \RH($T$), the in-plane resistivity $\rho_{ab}$($T$) and magnetoresistance (MR)
$\Delta\rho_{ab}/\rho_{ab}$($T$) are successfully modelled with parameters set solely by $\rho_{ab}$($T$). The
variation in $\Gamma_0$($\varphi$) is consistent with that proposed by Abrahams and Varma (AV) \cite{Abrahams00} to
explain the form of the single-particle scattering rate in cuprates \cite{Valla00}. AV attributed this form to
small-angle scattering off dopant atoms located between the CuO$_2$ planes \cite{Abrahams00, Varma01}. Our analysis
implies that such small-angle scattering may also appear in the {\it transport} lifetime and as such, play a crucial
role in determining the transport properties of cuprates.

Two bar-shaped samples (typical dimensions 1.5 x 0.5 x 0.1mm$^3$) were prepared from a large single crystal of LSCO30
grown in an infrared image furnace. Three pairs of 25 $\mu$m gold wires were contacted to each sample using silver
paint in a standard Hall configuration. The absolute error in $\rho_{ab}(T)$ due to uncertainty in the sample
dimensions is $\simeq 20\%$, as shown in Fig.~3a. \RH~and \MRa~were measured (1.4K $<T<$ 300K) using an ac lock-in
technique with low-noise transformers in a superconducting magnet with {\bf H}$\parallel$$c$. The Hall coefficient was
extracted using $R_{\rm H}=V_{\rm H}t/\mu_0IH$ where $V_{\rm H}=[V_{\rm H}$(10T)$-V_{\rm H}$(-10T)]/2 and $t$ is the
sample thickness. The orbital transverse MR was obtained by subtracting the positive linear MR measured in the
longitudinal configuration {\bf H}$\parallel${\bf I} at each temperature.

\begin{figure}
\includegraphics[width=7.0cm,keepaspectratio=true]{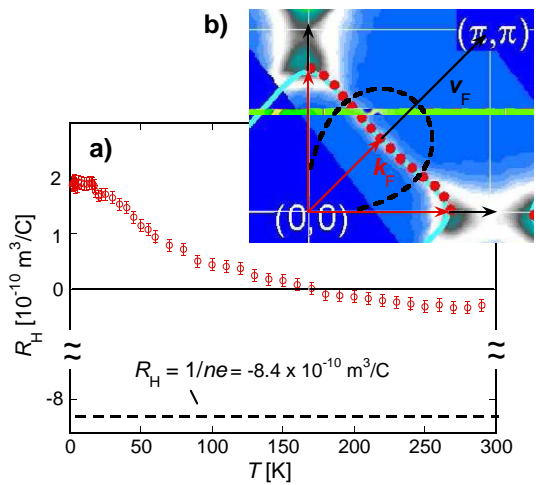}
\caption{({\bf a}) \RH($T$) for \LSCOOD; the Drude value (1/$ne$) is shown as a dashed line. {\bf b}) Fermi surface of
\LSCOOD~(red dots) \cite{Yoshida06} and tight-binding parameterization of the band structure (blue curve) (see
Table~\ref{BandPar}). The black dashed line represents the corresponding Fermi velocity whose peak value at $\varphi =
\pi/4$ is $v_F(\varphi)$ = 4$\times$10$^5$ ms$^{-1}$.} \label{Fig1}
\end{figure}

The measured $R_{\rm H}$, displayed in Fig.1a, shows the typical behavior observed at this doping level \cite{Hwang94,
Tsukada06}. At high $T$, $R_{\rm H}$ attains a constant value of $\sim -0.4\times10^{-10}$m$^{3}$/C. As $T$ is lowered,
$R_{\rm H}$ changes sign, saturating to a positive value below 15K of $\sim +2\times10^{-10}$m$^{3}$/C. The FS of
LSCO30, as determined by ARPES \cite{Yoshida06}, is reproduced in Fig.1b (red dots). A large, single, diamond-shaped FS
is seen, centered at (0, 0). The electrical and thermal properties of LSCO30 are characteristic of a correlated
Fermi-liquid: the Wiedemann-Franz law is obeyed \cite{Nakamae03}, $\rho_{ab}$($T$) = $\rho_0$ + $AT^2$ below 50K
\cite{Nakamae03}, and though elevated, the Kadowaki-Woods ratio ($A$/$\gamma_0^2$, where $\gamma_0$ is the electronic
specific heat coefficient) is consistent with band structure \cite{Hussey05}. Hence one might also expect the Hall data
to be consistent with Boltzmann transport theory. Within the isotropic-$\ell$ approximation however, $R_{\rm H}$ =
1/$ne \sim$ -8.4 $\times 10^{-10}$m$^3$/C (dashed line in Fig.1a), almost 20 times larger than $R_{\rm H}$(300) and of
opposite sign to $R_{\rm H}$(0).

One important aspect of the FS topology shown in Fig. 1b (and not resolved in earlier measurements \cite{Ino02}) is the
slight negative curvature as one moves from one apex of the diamond to another. This gives rise to alternating sectors
on the FS that have electron- and hole-like character. Ong showed that for a 2D metal in the weak-field semiclassical
limit, $\sigma_{xy}$ is determined by the \lq Stokes' area $\mathcal{A}$ (= $\int {\rm d}\ell$({\bf k}) $\times$
$\ell$({\bf k})) swept out by $\ell$({\bf k}) as {\bf k} moves around the FS. The local curvature of the FS gives rise
to different \lq circulation' of the $\ell$-vector and hence a contribution to $\sigma_{xy}$ with opposing sign. If,
{\it and only if}, $\ell$ is significantly different on the different parts of the FS, can an overall sign-change in
$\sigma_{xy}$ (and hence \RH) occur.

\begin{figure}
\includegraphics[width=7.5cm,keepaspectratio=true]{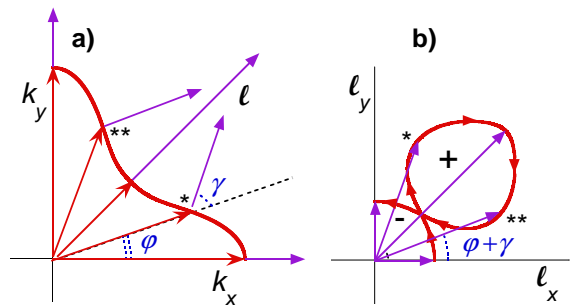}
\caption{({\bf a}) Section of 2D Fermi surface with pronounced negative curvature. The purple (red) arrows indicate the
direction and length of $\ell$($\varphi$) ($k_F(\varphi$)), as explained in the text. ({\bf b}) Polar plot of
$\ell$$(\varphi +\gamma)$. The red arrows indicate the circulation of each loop and the -/+ signs indicate the
corresponding sign of $\sigma_{xy}$. The resultant $\sigma_{xy}$ is determined by the difference in the areas of the
two counter-rotating loops ($\times 4$) \cite{Ong91}.} \label{Fig2}
\end{figure}

This effect is illustrated schematically in Fig.~\ref{Fig2}. The solid red line in Fig.~2a represents a 2D-projected FS
with exaggerated negative curvature. The purple arrows indicate the direction and length of the ${\bf \ell}$-vector for
selected points on the FS. The angles between ${\bf \ell}$ and {\bf k} and between {\bf k} and the $k_x$ axis are
labelled $\gamma$ and $\varphi$ respectively. As {\bf k} moves along the FS away from the $k_x$ axis, $\varphi$ and
$\gamma$ increase in the same sense. At a particular {\bf k}-point, marked by $*$, $\kappa$ = d$\gamma$/d$\varphi$
changes sign and remains negative until $\varphi$ reaches $**$. At $\varphi = \pi/2$, $\gamma$ is once again equal to
zero. If ${\bf \ell}$ is anisotropic, as shown in Fig.~2b, loops of different circulation will appear in the
$\ell_x$-$\ell_y$ plane. Ong demonstrated that $\sigma_{xy}$ will be determined by the sum of the areas enclosed by the
primary (negative) and secondary (positive contribution to $\sigma_{xy}$) loops.

Let us now examine quantitatively the situation in LSCO30. The Fermi wave vector $k_F(\varphi)$ (blue curve in Fig.1b)
and Fermi velocity ($\hbar \bf{v}_{F}=\bf{\nabla_k}\mathcal{E}_{k}$) (black dashed line in Fig.1b) are obtained from
the tight-binding expression used by Yoshida {\it et al.} to map the FS underlying the quasiparticle peaks of their
ARPES measurements on LSCO30~\cite{Yoshida06}: $\mathcal{E}_{k}=\mathcal{E}_{0}-2t$(cos$k_{x}a+$cos$k_{y}a)
-4t^{\prime}$(cos$k_{x}a$)(cos$k_{y}a) -2t^{\prime\prime}$(cos 2$k_{x}a+$cos$2 k_{y}a)$, where $t$, $t^{\prime}$ and
$t^{\prime\prime}$ are the first, second and third nearest neighbor hopping integrals between Cu sites respectively.
Note in Fig. 1b that the anisotropy in $v_F(\varphi)$ is large ($v_F(\pi/4)/v_F(0)=3.5$) even at this elevated doping
level.

To calculate $\rho_{ab}$, \RH~and \MRa, we invert the conductivity tensor $\sigma_{ij}$ for a quasi-2D metal following
Ref. \cite{Hussey03} where $\sigma_{ij}$ is given in terms of $k_F(\varphi)$, $v_F(\varphi)$ and $\Gamma(\varphi)$.
Assuming $\Gamma$ is isotropic at high $T$, one obtains \RH(300) = -1.0 $\times 10^{-10}$ m$^3$/C directly from the
band parameters given in Table~\ref{BandPar}, more than 8 times smaller than the isotropic-$\ell$ limit. Thus, the
ARPES-derived band anisotropy can account for almost all ($> 90\%$) of the deviation of \RH~from the classical Drude
result.

\begin{figure}
\includegraphics[width=7.0cm,keepaspectratio=true]{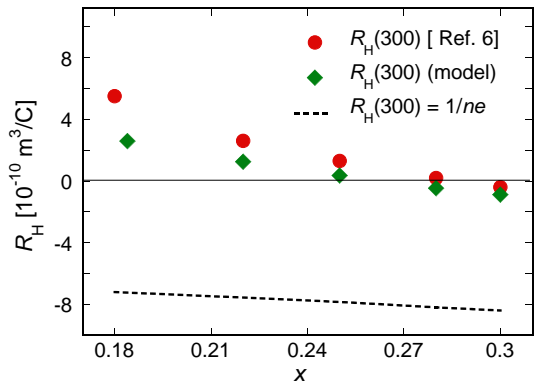}
\caption{Measured (red circles, \cite{Hwang94}) and band-derived (green diamonds, Table~\ref{BandPar}) \RH(300) values
in overdoped LSCO compared to the Drude value 1/$ne$ (dashed line).} \label{Fig3}
\end{figure}

Encouraged, we proceed to consider other values of $x$. According to ARPES, $\epsilon_F$ crosses the van Hove
singularity near $x \sim 0.18$, at which point, the FS becomes centered around ($\pi, \pi$) \cite{Yoshida06}. As the FS
approaches the saddle point, $t'/t$ increases (see Table~\ref{BandPar}), the FS curvature is enhanced and the
anisotropy in $v_F(\varphi)$ increases, implying even stronger renormalization of \RH(300) with decreasing $x$. To
illustrate this, we show in Fig.~\ref{Fig3} the experimentally determined values of \RH(300) for different $x \geq$
0.18 \cite{Hwang94} together with \RH~calculated using just the band parameters extracted from ARPES (see
Table~\ref{BandPar}).

The experimental trend is well captured, both qualitatively and quantitatively, by the band parameterization. By
contrast, the expected Drude result (dashed line in Fig.~\ref{Fig3}) remains large and negative at all $x$. This
appears to confirm that band anisotropy strongly renormalizes \RH(300) across the doping range. In order to account for
any discrepancy, only small changes in the FS topology (principally in $t'/t$), the inclusion of vertex corrections, or
some degree of three dimensionality, believed to develop in overdoped LSCO \cite{Kimura96, Hussey98} are required (as
$c$-axis warping is stronger at the saddles \cite{Yoshida06, Sahrakorpi05}, the effective {\it in-plane} $v_F$ at
($\pi, 0$) will be reduced, thus enhancing the overall anisotropy in $v_F$($\varphi$)). Fine details notwithstanding,
we believe that the transparency of the present analysis provides strong evidence that the ARPES-derived FS of LSCO is
essentially correct and can successfully account for the doping evolution of \RH(300) of overdoped LSCO without the
need for additional physics. This is significant as it is widely believed that the physical properties of LSCO depart
substantially from those of a conventional band picture as one approaches optimal doping. At lower doping ($x < 0.18$),
additional factors, such as the reduced Fermi \lq arc' length \cite{Yoshida03} or the development of a charge transfer
gap \cite{Ono07}, will be required to account for the full $T$- and $x$-dependence of \RH.

\begin{table}
\begin{center}
\begin{ruledtabular}
\begin{tabular}{ccccc}
  $x$  & $t$ (eV) & $\mathcal{E}_{0}/t$ & $-t^{\prime}/t$ & $-t^{\prime\prime}/t^{\prime}$\\
  \hline
 0.30 & 0.25 & 0.990 & 0.120 & 0.5\\
 0.28 & 0.25 & 0.960 & 0.121 & 0.5\\
 0.25 & 0.25 & 0.918 & 0.125 & 0.5\\
 0.22 & 0.25 & 0.880 & 0.130 & 0.5\\
 0.18 & 0.25 & 0.837 & 0.140 & 0.5\\
 \end{tabular}
\caption{\label{BandPar} Tight-binding band parameters for LSCO (0.18 $\leq x \leq$ 0.3) interpolated from values given
in ~\cite{Yoshida06}.}
\end{ruledtabular}
\end{center}
\end{table}


For all $x$ shown in Fig.~\ref{Fig3}, \RH(0) $>$ \RH(300) \cite{Hwang94}, implying that additional anisotropy develops
as $T \rightarrow 0$. In the absence of any experimental evidence for $T$-dependent FS reconstruction in LSCO, we
deduce that this additional anisotropy must be contained in the {\it elastic} scattering rate $\Gamma_0(\varphi)$,
presumably due to static impurities. According to AV, anisotropy in $\Gamma_0(\varphi)$ can arise from small-angle
scattering off dopant impurities located between the CuO$_2$ planes \cite{Abrahams00}. If $d$ is the characteristic
distance of such dopants from a plane, the electron scattering will involve only small momentum transfers $\delta k
\leq d^{-1}$. Then, $\Gamma_0$($\varphi$) is proportional to $\delta k$ and the local density of states, i.e. to
1/$v_F(\varphi)$. AV applied the same form of $\Gamma_0(\varphi)$ to the transport lifetime in order to interpret the
$T$-dependence of the inverse Hall angle in cuprates within a marginal Fermi-liquid framework \cite{Varma01}. Although
their derivation has subsequently been criticized \cite{Yakovenko, Carter02, Varma03}, a predominance of forward
impurity scattering in cuprates has been invoked to explain the weak suppression of $T_c$ with disorder {\cite{Kee01},
and the energy and $T$-dependence of the single-particle scattering rate $\Sigma''$ below $T_c$ \cite{Zhu04}.

In order to fit the transport data, we consider two functional forms of $\Gamma_0(\varphi)$: the AV form,
$\Gamma_0(\varphi)$ = $\beta/v_F(\varphi)$, and a squared sinusoid, $\Gamma_0(\varphi) = G_0(1 +
\chi$cos$^{2}(2\phi))$, which makes $\Gamma_0$ less peaked at $\varphi = 0, \pi/2$. For LSCO30, $\rho_{ab}$($T$) =
$\rho_0$ + $AT^2$ below 50K \cite{Nakamae03}. We assume, as found in overdoped Tl$_2$Ba$_2$CuO$_{6+\delta}$ (Tl2201)
\cite{Majed06}, that the $T^2$ scattering rate is isotropic within the basal plane, i.e. it is the {\it inelastic}
scattering that causes \RH($T$) to drop with increasing $T$. The intrinsic transport scattering rate can thus be
written as $\Gamma(\varphi,T) = \Gamma_0(\varphi) + \alpha T^2$. For the AV fit, the parameters $\alpha$ (=
1.6$\times$10$^9$ s$^{-1}$K$^2$) and $\beta$ (= 4.0$\times$10$^{18}$ ms$^{-2}$) are constrained by $\rho_{ab}(T)$,
implying there are no free parameters in the fitting of \RH($T$) and \MRa($T$). Note that this is not the case for the
sinusoidal function. Finally a high-$T$ saturation component to the scattering rate (again parameter-free) is
introduced, $\Gamma_{\rm max}=\langle {\bf v}_{\rm F} \rangle/a$, consistent with the Ioffe-Regel limit ($a$ being the
lattice parameter) \cite{Hussey03, Hussey04}, to take into account the deviation from $T^2$ resistivity above 50K and
to ensure that $\Gamma$ eventually becomes isotropic. This leads to an effective scattering rate $\Gamma_{\rm
eff}^{-1}(\varphi,T) = \Gamma^{-1}(\varphi,T) + \Gamma_{\rm max}^{-1}$ that is input into the calculation of
$\sigma_{ij}$.

\begin{figure}
\includegraphics[width=7.0cm,keepaspectratio=true]{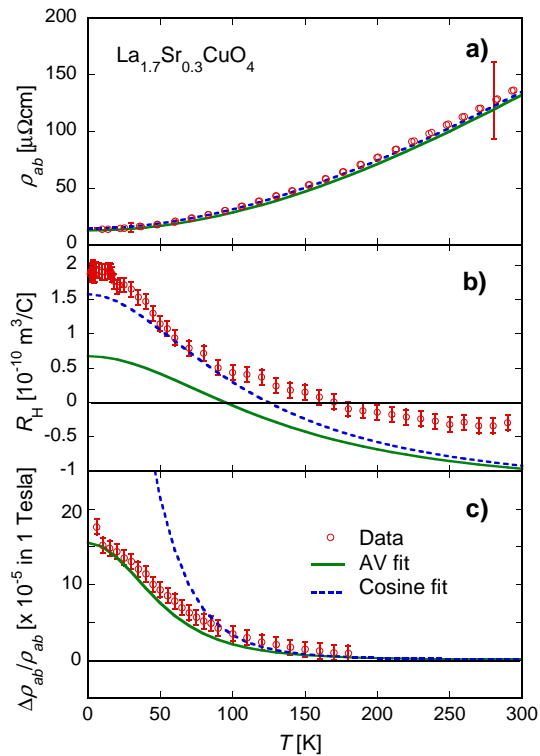}
\caption{{\bf a}) $\rho_{ab}$($T$), {\bf b}) $R_{H}$($T$) and {\bf  c}) \MRa($T$) (MR) data (red open circles) for
La$_{1.7}$Sr$_{0.3}$CuO$_4$. The green solid (blue dotted) lines are the fits using the AV (sinusoidal) form for
$\Gamma_0(\varphi)$ respectively (see text for details). The error bars in {\bf a}) give the level of uncertainty in
the sample dimensions, whilst in {\bf b}) and {\bf c}) they reflect scatter in the field sweeps.} \label{Fig4}
\end{figure}

The experimental results and model outputs for $\rho_{ab}$, \RH~and \MRa~are compared in Fig.~\ref{Fig3}. All AV fits
are in good quantitative agreement with the data at all $T$, despite there being no free parameters. The fit to
\RH($T$) using the sinusoidal form of $\Gamma_0(\varphi)$ (dotted line in Fig.~4b)) is equally good. (Here, $G_0$ = 7.4
$\times$ 10$^{12}$ s$^{-1}$ and $\chi$ = 3.3). However the same parameterization also leads to an overestimate of
\MRa~by a factor of 6 at low $T$. Indeed, one cannot fit both \RH~and \MRa~satisfactorily using this form for
$\Gamma_0(\varphi)$, despite the greater flexibility in the parameterization. This shortcoming demonstrates the
sensitivity of the MR (a second-order process) to the form of the anisotropy in $\Gamma_0(\varphi)$ and suggests that
the AV form is the more appropriate here. It is interesting to note that in La-doped Sr$_2$RuO$_4$, a very small amount
of off-plane Sr-site substitution ($\sim 1\%$) induces a sign change in \RH(0) even though the FS topology, as revealed
by quantum oscillations, is unchanged \cite{Kikugawa04}. Just as in LSCO, Sr$_2$RuO$_4$ contains both electron- and
hole-like regions of FS. This supports our conjecture that it is anisotropy in $\Gamma_0(\varphi)$, rather than FS
reconstruction, that causes \RH($T$) in LSCO to grow with decreasing $T$.

Finally, we note that in Tl2201, the isotropic-$\ell$ approximation appears to hold at low $T$ \cite{Hussey03b,
Majed06}, even in the presence of significant out-of-plane disorder (both interstitial oxygen and Cu substitution on
the Tl site). The key difference here is perhaps the band anisotropy. According to ARPES \cite{Plate05}, $v_F(\varphi)$
varies by $<$ 50$\%$ within the basal-plane \cite{Analytis07}, compared with $> 300\%$ in LSCO30. Hence, even if the
small-angle scattering were dominant in Tl2201, it would have a much smaller effect on \RH(0).

In summary, the Hall coefficient in overdoped LSCO is found to be extremely sensitive to details in the band structure
and that in the absence of a change in carrier number or a FS modulation, interpretation of \RH($T$) requires the
inclusion of strong in-plane anisotropy in the elastic scattering rate, presumably caused by small-angle scattering off
out-of-plane Sr substitutional disorder. Very recently, strong (factor of 3 or more) variation in $\Sigma''(\varphi)$
{\it of the AV symmetry} was reported in LSCO over the entire doping range \cite{Yoshida07}. Our analysis of \RH($T,
x$) affirms that this same form is manifest in the transport lifetime too. This striking violation of the
isotropic-$\ell$ approximation, which may apply not only to the cuprates, but also to other 2D correlated systems,
makes interpretation of \RH~in terms of carrier number highly inappropriate and misleading. It would certainly be
instructive to learn what effect this will have on future interpretation of \RH($T,x$) in the underdoped regime.

We thank K. Behnia, S. Carr, A. V. Chubukov, J. R. Cooper, R. A. Cooper, J. D. Fletcher, L. Pascut, A. J. Schofield, J.
A. Wilson and T. Yoshida for help and assistance. This work was supported by the EPSRC (U.K.).


\end{document}